\documentclass[10pt,conference]{IEEEtran}

\IEEEoverridecommandlockouts

\usepackage{amsmath}

\usepackage{color}
\usepackage[normalem]{ulem}

\usepackage{url}
\usepackage{comment}
\usepackage{subcaption}
\usepackage{xspace}
\usepackage{graphicx}
\usepackage[hidelinks,citecolor=blue,urlcolor=blue]{hyperref}
\usepackage[]{breakurl}
\usepackage{listings}
\usepackage[scaled=.8]{beramono}

\usepackage{algorithmic}
\usepackage{algorithm}
\usepackage[T1]{fontenc}
\usepackage{longtable}
\usepackage{pifont}
\usepackage{tikz}
\usetikzlibrary{arrows}

\begin{document}

\setlength{\pdfpageheight}{\paperheight}
\setlength{\pdfpagewidth}{\paperwidth}

% \conferenceinfo{SELSE '17}{March 21--22, 2017, Boston, MA, USA}
%\copyrightyear{20yy}
%\copyrightdata{978-1-nnnn-nnnn-n/yy/mm}
%\doi{nnnnnnn.nnnnnnn}

%\titlebanner{banner above paper title}        % These are ignored unless
%\preprintfooter{short description of paper}   % 'preprint' option specified.

\title{\Large\bf Adapting the DMTCP Plugin Model for Checkpointing
	of Hardware Emulation\footnotemark$^\dag$}

\author{
\IEEEauthorblockN{Rohan Garg$^*$\thanks{$^*$ This work was partially supported
           by the National Science Foundation under Grant~ACI-1440788.}}
\IEEEauthorblockA{Northeastern University\\
  Boston, MA\\
  \small{Email: rohgarg@ccs.neu.edu}}
\and
\IEEEauthorblockN{Kapil Arya}
\IEEEauthorblockA{Mesosphere, Inc.\\
  San Francisco, CA\\
  \small{Email: kapil@mesosphere.io}}
\and
\IEEEauthorblockN{Jiajun Cao$^*$}
\IEEEauthorblockA{Northeastern University\\
  Boston, MA\\
  \small{Email: jiajun@ccs.neu.edu}}
\and
\IEEEauthorblockN{Gene Cooperman$^*$}
\IEEEauthorblockA{Northeastern University\\
  Boston, MA\\
  \small{Email: gene@ccs.neu.edu}}
\and
\IEEEauthorblockN{Jeff Evans}
\IEEEauthorblockA{Mentor Graphics Corp.\\
  Austin, TX\\
  \small{Email: jeff\_evans@mentor.com}}
\and
\IEEEauthorblockN{Ankit Garg}
\IEEEauthorblockA{Mentor Graphics Corp.\\
  NOIDA / India\\
  \small{Email: Ankit\_Garg@mentor.com}}
\and
\IEEEauthorblockN{Neil A. Rosenberg}
\IEEEauthorblockA{Intel Corporation\\
  Austin, TX\\
\small{Email: neil.a.rosenberg@intel.com}}
\and
\IEEEauthorblockN{K. Suresh}
\IEEEauthorblockA{Mentor Graphics Corp.\\
  NOIDA / India\\
  \small{Email: K\_Suresh@mentor.com}}
}

%\authorinfo{}{}{}
\maketitle
{\let\thefootnote\relax\footnotetext{$^\dag$SELSE '17, March 21--22, 2017, Boston, MA, USA}}

% Use the following at camera-ready time to suppress page numbers.
% Comment it out when you first submit the paper for review.
% \thispagestyle{empty}

\begin{abstract}
Checkpoint-restart is now a mature technology.  It allows a user to
save and later restore the state of a running process.  The new plugin
model for the upcoming version~3.0 of DMTCP (Distributed MultiThreaded
Checkpointing) is described here.  This plugin model allows a target
application to disconnect from the hardware emulator at checkpoint time
and then re-connect to a possibly different hardware emulator at the
time of restart.
The DMTCP plugin model is important in allowing three distinct parties
to seamlessly inter-operate.  The three parties are: the EDA designer,
who is concerned with formal verification of a circuit design; the DMTCP
developers, who are concerned with providing transparent checkpointing
during the circuit emulation; and the hardware emulator vendor, who provides
a plugin library that responds to checkpoint, restart, and other events.

The new plugin model is an example of process-level virtualization:
virtualization of external abstractions from within a process. This
capability is motivated by scenarios for testing circuit models with the
help of a hardware emulator.  The plugin model enables a three-way
collaboration:  allowing a circuit designer and emulator vendor to each
contribute separate proprietary plugins while sharing an open source
software framework from the DMTCP developers.  This provides
a more flexible platform, where different fault injection models based
on plugins can be designed within the DMTCP checkpointing framework.
After initialization, one restarts from a checkpointed state under the
control of the desired plugin.  This restart saves the time spent in
simulating the initialization phase, while enabling fault injection
exactly at the region of interest.  Upon restart, one can inject faults
or otherwise modify the remainder of the simulation.  The work concludes
with a brief survey of the existing approaches to checkpointing and to
process-level virtualization.
\end{abstract}

%%%%%%%%%%%%%%%%%%%%%%%%%%%%%%%%%%%%%%%%%%%%%%%%%%%%%%%%%%%%%%%%%%%%%%
%%%%%%%%%%%%%%%%%%%%%%%%%%%%%%%%%%%%%%%%%%%%%%%%%%%%%%%%%%%%%%%%%%%%%%
\section{Introduction}

Checkpoint-restart is now a mature technology with a variety of
robust packages~\cite{ansel2009dmtcp,BLCR06,criu}.  This work concentrates
on the DMTCP (Distributed MultiThreaded CheckPointing) package
and its sophisticated plugin model that enables
process virtualization~\cite{arya2016design}.
This plugin model has been used recently to demonstrate
checkpointing of 32,752 MPI processes on a supercomputer
at TACC (Texas Advanced Computing Center)~\cite{cao2016system}.
DMTCP itself is free and open source.  The DMTCP publications
page~\cite{dmtcpPublications} lists approximately 50~refereed publications
by external groups that have used DMTCP in their work.

This work concentrates on the recent advances in the DMTCP programming model
that were motivated by work with Intel Corporation.  While Intel
works with multiple vendors of hardware emulators, this work
reflects the three-way collaboration between the DMTCP team,
Intel, and Mentor Graphics, a vendor of hardware emulators for EDA.
Further information specific to EDA (Electronic Design Automation)
is contained in~\cite{dac2017}.
In particular, the ability to save the state of a simulation
{\em including the state of a back-end hardware emulator} is a key to using
checkpoint-restart in EDA.

For background
on how DMTCP is used generally at Intel, see~\cite{hpec2014}.
The focuses of the ongoing work at Intel is best described
by their statement of future work:
\begin{quotation}
\noindent
``Within Intel IT, we will focus on the development and
enhancement of the DMTCP technology for use with
graphical EDA tools, with strong network dependencies.
$\ldots$
There is also additional engagement with third-party vendors
to include native DMTCP support in their tools, as well as
engagement with super-computing development teams on
enabling DMTCP for the Xeon Phi family of products.''
\end{quotation}

A hardware emulator may entail a thousand-fold
slowdown, as compared to direct execution in silicon.
There are two natural use cases of checkpointing in the context of EDA.
In both cases, the natural strategy is to run until reaching
the region of logic of interest.  Then checkpoint.  Later,
one can repeatedly restart and test the logic, without worrying
about the long initialization times under a hardware emulator.
Restarting under DMTCP is extremely fast, especially when the
{\tt --fast-restart} flag is used that takes advantage of {\tt mmap()}
to load pages into memory on-demand at runtime (after the initial restart).
The two use cases follow.
\begin{LaTeXdescription}
  \item[Testing of silicon logic:] run until reaching the logic to
	be tested; then repeatedly restart and follow different
	logic branches; and
  \item[Fault injection in silicon logic:] run until reaching the logic
	to be tested; then repeatedly restart, inject faults in the
	emulated (or simulated) silicon model and run along a pre-determined
	logic branch to determine the level of fault tolerance for that
	silicon design.
\end{LaTeXdescription}
For this work, the second case is of greater interest.  This requires
running arbitrary code either immediately at the point of restart by
injecting faults in the logic design,
or by interposing on later logic functions of the simulator/emulator
so as to inject transient faults.

The first use case above has been extensively studied
using DMTCP
in domains as varied as architecture simulation~\cite{ShinaEtAl12},
formal verification of embedded control
systems~\cite{resmerita2012verification},
network simulation~\cite{HarriganRiley14},
and software model checking~\cite{LeungwattanakitEtAl14}.  While the
two use cases are closely related, this work
highlights the second use case, by including the possibility of
interposing at runtime.  Section~\ref{sec:processVirtualization}
presents the tools for such interposition, including the creation
of global barriers at an arbitrary point in the program.
Section~\ref{sec:caseStudies} presents three particular extensions
of checkpointing that were added to the DMTCP plugin model specifically
motivated by the concerns observed in our general collaboration on EDA.

The DMTCP plugin model is critical in this latter application.
One must stop a computation at a pre-defined location in the
simulation, save additional state information (such as the
state of a hardware emulator being used~\cite{dac2017}), and
then inject additional code (such as fault injection) at restart time.
A contribution of the DMTCP plugin model is the ability to virtualize
multiple aspects of the computation.  These include:  pathnames (for example,
the subdirectory corresponding to the current ``run slot'' of the
emulator);
environment variables (for example, modification of
the DISPLAY environment variable, or other environment variables
intrinsic to the running of the simulation);
interposition of the simulation by a third-party plugin (for example,
for purposes of measuring timings since restart at multiple levels
of granularity, or programmatically creating additional checkpoints
for analysis of interesting states leading to logic errors);
and third-party programmable barriers across all processes (enabling
the acceleration of simulations through the use of parallel
processes and even distributed processes within a single computation).

The DMTCP plugin model is an example of {\em process virtualization}:
virtualization of external abstractions from within a process.
It is argued here that the DMTCP plugin model sets it apart from
other checkpointing approaches.  To this end, a brief survey
of existing checkpointing approaches and process virtualization
is provided at the end.

In the rest of this paper, Section~\ref{sec:processVirtualization}
motivates the need for a model of process virtualization with a simple
example concerning process ids.  It also reviews the DMTCP plugin model.
Section~\ref{sec:caseStudies} presents a series of micro-case studies
in which DMTCP was extended to support the applications at Intel,
along with third-party DMTCP plugins developed by Mentor Graphics
for use by Intel and other customers..
Section~\ref{sec:relatedWork} the provides a survey of DMTCP
and some other related approaches to checkpointing and process
virtualization.
Section~\ref{sec:conclusion} then presents the conclusions.

%%%%%%%%%%%%%%%%%%%%%%%%%%%%%%%%%%%%%%%%%%%%%%%%%%%%%%%%%%%%%%%%%%%%%%
%%%%%%%%%%%%%%%%%%%%%%%%%%%%%%%%%%%%%%%%%%%%%%%%%%%%%%%%%%%%%%%%%%%%%%
\section{User-Space Process Virtualization}
\label{sec:processVirtualization}

Application-specific checkpointing and system-level transparent checkpointing
are two well-known options for checkpointing.  Unfortunately, neither
one fits the requirements for the proposed use case for simulating
fault injection in silicon logic.  Application-specific checkpointing
is error-prone and difficult to maintain.  System-level transparent
checkpointing generally does not provide enough control at runtime
to dynamically adjust the type of fault injection.
In particular, it is often necessary to
capture control of the target application dynamically at runtime in
order to inject faults.  Here we show how that can be incorporated
in a modular DMTCP plugin, rather than incorporated directly into
the simulator/emulator.

For a more thorough introduction to the DMTCP plugin model, see
either~\cite{arya2016design} or the DMTCP documentation~\cite{dmtcpPlugins}.
This section highlights those aspects most likely to assist in adding
fault injection through a DMTCP plugin.

The primary features of the model of interest for fault injection are:
\begin{enumerate}
  \item interposition on function/library calls, and their use
	in virtualization;
  \item programmatically defined barriers across all processes on
	a computer; and
  \item programmatically defined choices of when to checkpoint and
	when to avoid checkpointing.
\end{enumerate}

%%%%%%%%%%%%%%%%%%%%%%%%%%%%%%%%%%%%%%%%%%%%%%%%%%%%%%%%%%%%%%%%%%%%%%
\subsection{Process Virtualization through Interposition and Layers:
	  A Simple Example with Pids}
\label{sec:pidVirtExample}

\begin{figure}[ht]
  \begin{center}
    \includegraphics[width=0.9\columnwidth]{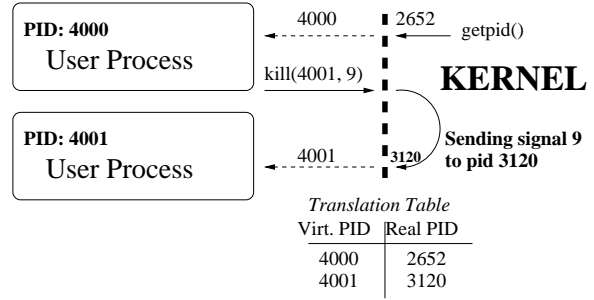}
  \end{center}
  \caption{Process virtualization for pids.}
  \label{fig:pid-virt}
\end{figure}

Figure~\ref{fig:pid-virt} succintly describes the philosophy of
process virtualization.  Some invariant (in this case the pid (process~id)
of a process may have a different name prior to checkpoint and after
restart.  A virtualized process will interact only with virtual
process ids in the base code.  A DMTCP plugin retains a translation
table between the virtualized pid known to the base code and
the real pid known to the kernel.

Since the base code and the kernel interact primarily through system calls,
the DMTCP plugin defines a wrapper function around that system call.
The wrapper function translates between virtual and real pids both
for arguments to the system call and for the return value.  This is
illustrated both in Figure~\ref{fig:pid-virt} and in the example
code of Listing~\ref{lst:pidWrapperExample}.

\begin{lstlisting}[belowskip=-20pt, float=ht,
  escapechar=@, label={lst:pidWrapperExample},
  morekeywords={WRAPPER,enable_ckpt,disable_ckpt,REAL_},
  caption={A simplified function wrapper for pid virtualization}]
WRAPPER int kill(pid_t pid, int sig) {
  disable_ckpt();
  real_pid = virt_to_real(pid);
  int ret = REAL_@@kill(real_pid, sig);
  enable_ckpt();
  return ret;
}
\end{lstlisting}
\vskip10pt

Additionally, pid's may be passed as part of the proc filesystem,
and through other limited means.  To solve this, DMTCP implements
virtualization of filenames as well as pid names, and so the ``open''
system call will also be interposed upon to detect names such
as \texttt{/proc/PID/maps}.

In this way, a collection of wrapper functions can be collected
together within a DMTCP plugin library.  Such a library implements
a virtualization layer.  The ELF library standard implements
a library search order such that symbols are searched in order as follows:
\newline
\centerline{
  EXECUTABLE $\rightarrow$ LIB1 $\rightarrow$ LIB2 ... LIBC $\rightarrow$ KERNEL
}
\newline
where the symbol is finally replaced by a direct kernel call.

This sequence can also be viewed as a sequence of layers, consistent
with the common operating system implementation through layers.  A DMTCP
plugin for pids then presents a virtualization layer in which all higher
layers see only virtual pids, and all lower layers see only real pids.
This is analogous to an operating system design in which a higher layer
sees the disk as a filesystem, and a lower layer sees the disk as a
collection of disk blocks.
In a similar way, DMTCP provides layers to virtualize filenames,
environment variables and myriad other names.

In this way, an end
user can implement a fault injection plugin layer such that
all code below that layer sees injected faults, while higher layers
do not see the injected faults.  Additionally, such a layer can
be instrumented to gather information such as the cumulative number
of faults.

DMTCP also provides an API for the application or a plugin to
either request a checkpoint or to avoid a checkpoint.  Upon checkpoint,
each plugin is notified of a checkpoint barrier, and similarly upon
restart.  Thus, it is feasible to create successive checkpoints
available for restart or available as a snaphot for later forensics
on the cause of a later error.  Optimizations such as forked checkpointing
(fork a child and continue in the parent) are available in order to
take advantage of the kernel's copy-on-write in order to
make checkpointing/snapshotting extremely fast.

%%%%%%%%%%%%%%%%%%%%%%%%%%%%%%%%%%%%%%%%%%%%%%%%%%%%%%%%%%%%%%%%%%%%%%%%%%%%%%%%%
%%%%%%%%%%%%%%%%%%%%%%%%%%%%%%%%%%%%%%%%%%%%%%%%%%%%%%%%%%%%%%%%%%%%%%%%%%%%%%%%%
\subsection{Checkpointing Distributed Resources with the Help of Barriers}
\label{sec:checkpointingDistributedResources}

Checkpointing in a distributed application context requires coordination
between multiple processes at different virtualization layers. The use
of programmable barriers enables this coordination. In addition to
the checkpoint and restart events, each plugin (or virtualization layer)
can define its own set barriers and a callback to execute at a barrier. A
centralized DMTCP coordinator forces the application processes to execute
the barriers in sequence.

Further, a hardware resource, for example, the interface to a hardware
emulator, might be shared among multiple processes that share parent-child
relationships. To get a semantically equivalent state on restart, the barriers
can be used to elect a leader to save and restore the connection to the
hardware emulator on restart.

%%%%%%%%%%%%%%%%%%%%%%%%%%%%%%%%%%%%%%%%%%%%%%%%%%%%%%%%%%%%%%%%%%%%%%%%%%%%%%
%%%%%%%%%%%%%%%%%%%%%%%%%%%%%%%%%%%%%%%%%%%%%%%%%%%%%%%%%%%%%%%%%%%%%%%%%%%%%%

% Section title:  Implementation -> Programming Model
% As we know, it's dangerous to devote a lot of space to "implementation"
% And this is the first detailed description of the model.  -- Gene

\section{Case Studies along the Way to Extending DMTCP}
\label{sec:caseStudies}

This section describes three specific real-world use cases where DMTCP
was extended to support hardware emulation and simulation software. The
examples are motivated by our work with various hardware and EDA tool
vendors.

\subsection{External connections}

GUI-based simulation software presents a unique challenge in
checkpointing.  The front-end software communicates with an X server
via a socket. The X server runs in a privileged mode and outside of
checkpoint control. While the connection could be blacklisted for the
checkpointing, application's GUI context and state is part of the X server
and cannot be checkpointed.  The context does not exist at restart time
and needs to be restored. DMTCP was extended to transparently support
checkpointing of VNC~\cite{richardson1998virtual}
and XPRA~\cite{xpra}. The two tools allow
X command forwarding to a local X server that can be run under checkpoint
control. \cite{KazemiGC13} presents an alternate record-prune-replay
based approach using DMTCP to checkpoint GUI-based applications.

Authentication and license services is an important issue for protecting
the intellectual property of all the parties. Often, the authentication
protocols and software are proprietary and specific to a vendor. Further,
the licensing services are not run under checkpoint control, which makes
it difficult to get a ``complete'' checkpoint of the software. Extensions
were added to DMTCP to allow a vendor to hook into the checkpoint
and restart events and mark certain connections as ``external'' to
the computation. At checkpoint time, the connections marked external
are ignored by DMTCP and instead the responsibility of restoring
these connections is delegated to the vendor-specific extension. The
vendor-specific plugin also allows the application to check back with
the licensing service at restart time in order to not violate a licensing
agreement that restricts the number of simultaneous ``seats''.

\subsection{Virtualizing an application's environment}

The ability to migrate a process among the available resources is critical
for efficient utilization of hardware emulator resources. However, the
environment variables, the file paths, and the files that are saved as
part of a checkpoint image make such migrations challenging. We added
DMTCP extensions (plugins) to virtualize the environment and the file
paths.  This allows a process to be restarted on a different system
by changing the values and the paths. Another extension that we added
to DMTCP allows a user to explicitly control the checkpointing of
files used by their application at the granularity of a single file.

\subsection{Interfacing with hardware and closed-source, third-party libraries}

Hardware emulators communicate with the host software via high-speed
interfaces. Any in-flight transactions at checkpoint time can result
in the data being lost and inconsistent state on restart. Thus,
it is important to bring the system to a quiescent state and drain
the in-flight data on the buses before saving the state. Further,
checkpointing while the software is in a critical state (like holding
a lock on a bus) can lead to complications on restart. To help mitigate
such issues, DMTCP was extended to allow fine-grained programmatic control
over checkpointing. This enables the hardware/EDA tool vendor to tailor
the checkpointing for their specific requirements. In particular, it
allows a user to invoke checkpointing from within their code, disable
checkpointing for critical sections, or delay the resuming of user
threads until the system reaches a well-behaved state.

The software toolchain used for simulation and emulation is often put
together by integrating various third-party components. The components
may be closed-source and may use proprietary protocols for interfacing
with each other and the system. For example,  many software toolchains
rely on legacy 32-bit code that's difficult to port to 64-bits, and so,
support for mixed 32-/64-~bit processes was an important consideration.

Checkpointing while holding locks was another interesting issue.  While
the locks and their states are a part of the user-space memory (and hence,
a part of the checkpoint image), an application can also choose to use
an error-checking lock that disallows unlocking by a different thread
than the one that acquired it. On restart, when new thread ids would be
assigned by the system, the locks would become invalid and the unlock
call would fail. We extended DMTCP by adding wrapper functions for lock
acquisition and release functions to keep track of the state of locks. At
restart time, a lock's state is patched with the newer thread ids.

More generally, the problem described above is about the state that's
preserved when a resource is allocated at checkpoint time and needs to
be deallocated at restart time. While the restarted process inherits
its state from the checkpoint image, its environment (thread ids, in
the above case) might have changed on restart. An application author
with domain expertise can extend the DMTCP checkpointing framework to
recognize and virtualize these resources. The state could be a part of
the locks that are acquired by a custom thread-safe malloc library, or
the guard regions created by a library to guard against buffer overflows,
or the libraries that are loaded temporarily.

%%%%%%%%%%%%%%%%%%%%%%%%%%%%%%%%%%%%%%%%%%%%%%%%%%%%%%%%%%%%%%%%%%%%%%%%
%%%%%%%%%%%%%%%%%%%%%%%%%%%%%%%%%%%%%%%%%%%%%%%%%%%%%%%%%%%%%%%%%%%%%%%%
\section{Survey of Existing Approaches to Checkpointing
	and Process Virtualization}
\label{sec:relatedWork}

High performance computing (HPC) is the traditional domain in which
checkpoint-restart is heavily used.  It is used for the sake of fault
tolerance during a long computation, for example of days.  For a
survey of checkpoint-restart implementations in the context of high
performance computing, see Egwutuoha\hbox{et al.}~\cite{Egwutuoha2013}.
In the context of HPC, DMTCP and BLCR~\cite{BLCR06,BLCR03} are the most
widely used examples of transparent, system-level checkpoint-restart
parallel computing.  (A transparent checkpointing package is one
that does not modify the target application.)

\subsection{DMTCP}
\label{sec:relatedWorkDMTCP}
DMTCP (Distributed MultiThreaded CheckPointing) is a purely user-space
implementation.  In addition to being transparent, it also does not
require any kernel modules and its installation and execution does
not require root privilege or the use of special Linux capabilities.
It achieves its robustness by trying to stay as close to the POSIX
standard as possible in its API with the Linux kernel.

The first version of DMTCP was later described in~\cite{ansel2009dmtcp}.
That version did not provide the plugin model for process virtualization.
For example, virtualization of network addresses did not exist,
as well as a series of other constructs, such as
timers, session ids, System~V shared memory, and other features.
These features were added later due to the requirements of
high performance computing.  Eventually, the current procedure
for virtualizing process ids (see Section~\ref{sec:pidVirtExample}
was developed.  To the best of our knowledge, DMTCP is unique
in its approach toward process id virtualization.

Eventually, the plugin model was developed, initially for transparent
support of the InfiniBand network fabric~\cite{cao2014transparent}.
the current extension of that plugin model is described
in~\cite{arya2016design}.

Still later, the requirements for robust support of EDA in collaboration
with Intel led to the development of reduction of runtime overhead
graphic support using XPRA, path virtualization (for virtualization
of the runtime slot and associated directory of a run using
a hardware emulator, including different mount points on
the restart computer), virtualization of environment variables
including the X-Windows DISPLAY variable (for similar reasons), robustness
across a variety of older and newer Linux kernels and GNU libc versions,
mixed multi-architecture (32- and 64-bit) processes within a single
computation, low-overhead support for malloc-intensive programs,
re-connection of a socket to a license server on restart, and whitelist
and blacklist of special temporary files that many or may not be present
on the restart computer.

\subsection{BLCR}
\label{sec:relatedWorkBLCR}
BLCR supports only single-node standalone checkpointing.  In particular,
it does not support checkpointing of TCP sockets, InfiniBand
connections, open files, or SysV shared memory objects.

BLCR is often used in HPC clusters, where one has full control over
the choice of Linux kernel and other systems software.  Typically,
a Linux kernel is chosen that is compatible with BLCR, a BLCR
kernel module is installed, and when it is time to checkpoint,
it is the responsibility of an MPI checkpoint-restart service
to temporarily disconnected the MPI network layer, then checkpoint
locally on each node, and finally re-connect the MPI network
layer.

Note that BLCR is limited in what features it supports, notably
including a lack of support for sockets and System~V shared memory.
Quoting from the BLCR User's Guide:
\begin{quote}
``However, certain applications are not supported because they use
  resources not restored by BLCR: $\ldots$
    Applications which use sockets (regardless of address family). $\ldots$;
    Applications which use character or block devices (e.g. serial ports
      or raw partitions). $\ldots$;
    Applications which use System~V IPC mechanisms including shared memory,
    semaphores and message queues.''~\cite{BLCRshmem}
\end{quote}
The lack of BLCR support for shared memory also prevents its use
in OpenSHMEM~\cite{openshmem}.

\subsection{ZapC and CRUZ}
ZapC and CRUZ represent two other checkpointing approaches
that are not currently widely used.

ZapC~\cite{Laadan2005} and CRUZ~\cite{CRUZ2005} were earlier
efforts to support distributed checkpointing, by modifying the kernel to
inserting hooks into the network stack using netfilter to translate source
and destination addresses.  ZapC and CRUZ are no longer in active use.
They were designed to virtualize primarily two resources:
process ids and IP network addresses.  They did not support
SSH, InfiniBand, System~V IPC, or POSIX timers, all of which are commonly
used in modern software implementation.

\subsection{CRIU}
CRIU~\cite{criu} leverages Linux namespaces for transparently
checkpointing on a single host (often within a Linux container),
but lacks support for distributed computations.  Instead of directly
virtualizing the process id, CRIU relies on extending the kernel API
through a much larger proc filesystem and a greatly extended ``prctl''
system call.   For example, the ``PR\_SET\_MM'' has 13~additional parameters
that can be set (e.g., beginning end end of text, data, and stack).
In another example, CRIU relies on the ``CONFIG\_CHECKPOINT\_RESTORE''
kernel configuration to allow a process to directly modify the kernel's
choice of pid for the next process to be created~\cite{criuSetPid}.
In a general context, there is a danger that the desired pid to be
restored may already be occupied by another process, but CRIU is also
often used within a container where this restriction can be avoided.

Finally, CRIU has a more specialized plugin facility~\cite{criuPlugins}.
Some examples are:  ability to save and restore the contents of particular
files; and the means to save and restore pointers to external sockets,
external links, and mount points that are outside the filesystem namespace
of an LXC (Linux Container).  Recall that CRIU does not try to support
distributed computations.  Perhaps it is for this reason that CRIU did
not have the same pressure to develop a broader plugin system capable
of supporting generic external devices such as hardware emulators.

\subsection{Process Virtualization}
The term {\em process virtualization} was used in~\cite{ValleeEtAl05}.
That work discusses kernel-level support for such process virtualization,
while the current work emphasizes an entirely user-space approach
within unprivileged processes.
Related to process virtualization is the concept of a Library~OS,
exemplified by the Drawbridge Library~OS~\cite{PorterEtAl11} and
Exokernel~\cite{EnglerEtAl95}.  However, such systems are concerned
with providing {\em extended or modified} system services that are not
natively present in the underlying operating system kernel.

Both process-level virtualization and the Library~OS approach employ a
user-space approach (ideally with no modification to the application
executable, and no additional privileges required).  However, a Library~OS
is concerned with providing {\em extended or modified} system services
that are not natively present in the underlying operating system kernel.
Process virtualization is concerned with providing a semantically
equivalent system object using the {\em same} system service.  This need
arises when restarting from a checkpoint image, or when carrying out a
live process migration from one computer to another.  The target computer
host is assumed to provide the same system services as were available
on the original host.

Although process-level virtualization and a Library~OS both operate
in user space without special privileges, the goal of a Library~OS
is quite different.  A Library~OS modifies or extends the system
services provided by the operating system kernel.  For example,
Drawbridge~\cite{PorterEtAl11} presents a Windows~7 personality,
so as to run Windows~7 applications under newer versions of Windows.
Similarly, the original exokernel operating system~\cite{EnglerEtAl95}
provided additional operating system services beyond those of a small
underlying operating system kernel, and this was argued to often be more
efficient that a larger kernel directly providing those services.

\begin{comment}
The CIFTS project~\cite{GuptaEtAl09} provides a fault-tolerance
backplane (FTB), based on an interface specification that
allows libraries and runtime systems to jointly manage faults.
\end{comment}

%%%%%%%%%%%%%%%%%%%%%%%%%%%%%%%%%%%%%%%%%%%%%%%%%%%%%%%%%%%%%%%%%%%%%%%%
%%%%%%%%%%%%%%%%%%%%%%%%%%%%%%%%%%%%%%%%%%%%%%%%%%%%%%%%%%%%%%%%%%%%%%%%
\vskip -12pt
\section{Conclusion}
\label{sec:conclusion}

In order to develop a successful plugin model for checkpointing
in the context of EDA, one required modularity that enabled
the DMTCP team, Intel, and Mentor Graphics to each write their
own modular code.  Further, the Intel and Mentor Graphics DMTCP-based plugins
and other code were of necessity proprietary.  This work has shown
how the DMTCP plugin model can be used to provide a flexible
model enabling full cooperation, while avoiding the more extreme
roadmaps of either fully application-specific code or transparent,
system-level checkpointing with no knowledge of the proprietary
aspects of the Mentor Graphics hardware emulator.



\begin{thebibliography}{10}
\providecommand{\url}[1]{#1}
\csname url@samestyle\endcsname
\providecommand{\newblock}{\relax}
\providecommand{\bibinfo}[2]{#2}
\providecommand{\BIBentrySTDinterwordspacing}{\spaceskip=0pt\relax}
\providecommand{\BIBentryALTinterwordstretchfactor}{4}
\providecommand{\BIBentryALTinterwordspacing}{\spaceskip=\fontdimen2\font plus
\BIBentryALTinterwordstretchfactor\fontdimen3\font minus
  \fontdimen4\font\relax}
\providecommand{\BIBforeignlanguage}[2]{{%
\expandafter\ifx\csname l@#1\endcsname\relax
\typeout{** WARNING: IEEEtran.bst: No hyphenation pattern has been}%
\typeout{** loaded for the language `#1'. Using the pattern for}%
\typeout{** the default language instead.}%
\else
\language=\csname l@#1\endcsname
\fi
#2}}
\providecommand{\BIBdecl}{\relax}
\BIBdecl

\bibitem{ansel2009dmtcp}
J.~Ansel, K.~Arya, and G.~Cooperman, ``{DMTCP: Transparent Checkpointing for
  Cluster Computations and the Desktop},'' in \emph{IEEE Int. Symp. on Parallel
  and Distributed Processing (IPDPS)}.\hskip 1em plus 0.5em minus 0.4em\relax
  IEEE Press, 2009, pp. 1--12.

\bibitem{BLCR06}
P.~Hargrove and J.~Duell, ``{Berkeley Lab Checkpoint/Restart (BLCR) for Linux
  Clusters},'' \emph{Journal of Physics Conference Series}, vol.~46, pp.
  494--499, Sep. 2006.

\bibitem{criu}
{CRIU team}, ``{CRIU},'' accessed Jan., 2017, \url{http://criu.org/}.

\bibitem{arya2016design}
K.~Arya, R.~Garg, A.~Y. Polyakov, and G.~Cooperman, ``Design and implementation
  for checkpointing of distributed resources using process-level
  virtualization,'' in \emph{IEEE Int. Conf. on Cluster Computing
  (Cluster'16)}.\hskip 1em plus 0.5em minus 0.4em\relax IEEE Press, 2016, pp.
  402--412.

\bibitem{cao2016system}
J.~Cao, K.~Arya, R.~Garg, S.~Matott, D.~K. Panda, H.~Subramoni, J.~Vienne, and
  G.~Cooperman, ``System-level scalable checkpoint-restart for petascale
  computing,'' in \emph{22nd IEEE Int. Conf. on Parallel and Distributed
  Systems (ICPADS'16)}.\hskip 1em plus 0.5em minus 0.4em\relax IEEE Press,
  2016, also, technical report available as: arXiv preprint arXiv:1607.07995.

\bibitem{dmtcpPublications}
{DMTCP team}, ``{DMTCP} publications,'' accessed Jan., 2017,
  \url{http://dmtcp.sourceforge.net/publications.html}.

\bibitem{dac2017}
G.~Cooperman, J.~Evans, A.~Garg, R.~Garg, N.~A. Rosenberg, and K.~Suresh,
  ``Transparently checkpointing software test benches to improve productivity
  of {SoC} verification in an emulation environment,'' 2017, (submitted).

\bibitem{hpec2014}
\BIBentryALTinterwordspacing
I.~Ljubuncic, R.~Giri, A.~Rozenfeld, and A.~Goldis, ``Be kind, rewind ---
  checkpoint \& restore capability for improving reliability of large-scale
  semiconductor design,'' in \emph{2014 IEEE High Performance Extreme Computing
  Conference (HPEC-2014)}, 2014. [Online]. Available:
  \url{http://www.ieee-hpec.org/2014/CD/index_htm_files/FinalPapers/34.pdf}
\BIBentrySTDinterwordspacing

\bibitem{ShinaEtAl12}
A.~Shina, K.~Ootsu, T.~Ohkawa, T.~Yokota, and T.~Baba, ``Proposal of
  incremental software simulation for reduction of evaluation time,'' in
  \emph{2012 Third International Conference on Networking and Computing
  (ICNC)}, Dec 2012, pp. 311--315.

\bibitem{resmerita2012verification}
S.~Resmerita and W.~Pree, ``Verification of embedded control systems by
  simulation and program execution control,'' in \emph{2012 American Control
  Conference (ACC)}.\hskip 1em plus 0.5em minus 0.4em\relax IEEE, 2012, pp.
  3581--3586.

\bibitem{HarriganRiley14}
K.~Harrigan and G.~Riley, ``Simulation speedup of ns-3 using checkpoint and
  restore,'' in \emph{Proceedings of the 2014 Workshop on ns-3}.\hskip 1em plus
  0.5em minus 0.4em\relax ACM, 2014, p.~7.

\bibitem{LeungwattanakitEtAl14}
W.~Leungwattanakit, C.~Artho, M.~Hagiya, Y.~Tanabe, M.~Yamamoto, and
  K.~Takahashi, ``Modular software model checking for distributed systems,''
  \emph{IEEE Trans. on Software Engineering}, vol.~40, no.~5, pp. 483--501, May
  2014.

\bibitem{dmtcpPlugins}
{DMTCP team}, ``dmtcp/plugin-tutorial.pdf,''
  \url{http://github.com/dmtcp/dmtcp/blob/master/doc/plugin-tutorial.pdf},
  accessed Jan., 2017.

\bibitem{richardson1998virtual}
T.~Richardson, Q.~Stafford-Fraser, K.~R. Wood, and A.~Hopper, ``Virtual network
  computing,'' \emph{IEEE Internet Computing}, vol.~2, no.~1, pp. 33--38, 1998.

\bibitem{xpra}
{XPRA team}, ``{XPRA},'' accessed Jan., 2017, \url{http://xpra.org/}.

\bibitem{KazemiGC13}
\BIBentryALTinterwordspacing
S.~Kazemi, R.~Garg, and G.~Cooperman, ``Transparent checkpoint-restart for
  hardware-accelerated 3d graphics,'' \emph{CoRR}, vol. abs/1312.6650, 2013.
  [Online]. Available: \url{http://arxiv.org/abs/1312.6650}
\BIBentrySTDinterwordspacing

\bibitem{Egwutuoha2013}
I.~P. Egwutuoha, D.~Levy, B.~Selic, and S.~Chen, ``\BIBforeignlanguage{en}{A
  survey of fault tolerance mechanisms and checkpoint/restart implementations
  for high performance computing systems},'' \emph{\BIBforeignlanguage{en}{The
  Journal of Supercomputing}}, vol.~65, no.~3, pp. 1302--1326, Sep. 2013.

\bibitem{BLCR03}
J.~Duell, P.~Hargrove, and E.~Roman, ``{The Design and Implementation of
  Berkeley Lab's Linux Checkpoint/Restart (BLCR)},'' Lawrence Berkeley National
  Laboratory, Tech. Rep. LBNL-54941, 2003.

\bibitem{cao2014transparent}
J.~Cao, G.~Kerr, K.~Arya, and G.~Cooperman, ``{Transparent Checkpoint-Restart
  over {I}nfini{B}and},'' in \emph{Proc. of the 23rd Int. Symp. on
  High-performance Parallel and Distributed Computing}.\hskip 1em plus 0.5em
  minus 0.4em\relax ACM Press, 2014, pp. 13--24.

\bibitem{BLCRshmem}
{BLCR team}, ``Berkeley {L}ab {C}heckpoint/{R}estart ({BLCR}) user's guide,''
  accessed Jan., 2017,
  \url{https://upc-bugs.lbl.gov/blcr/doc/html/BLCR_Users_Guide.html}.

\bibitem{openshmem}
{OpenSHMEM team}, ``Openshmem,'' accessed Jan., 2017,
  \url{http://openshmem.org/site/Links#imp}.

\bibitem{Laadan2005}
O.~Laadan, D.~Phung, and J.~Nieh, ``Transparent checkpoint-restart of
  distributed applications on commodity clusters,'' in \emph{Cluster Computing,
  2005. {IEEE} International}, Sep. 2005, pp. 1--13.

\bibitem{CRUZ2005}
G.~Janakiraman, J.~Santos, D.~Subhraveti, and Y.~Turner, ``Cruz:
  Application-transparent distributed checkpoint-restart on standard operating
  systems,'' in \emph{International Conference on Dependable Systems and
  Networks, 2005. {DSN} 2005. Proceedings}, Jun. 2005, pp. 260--269.

\bibitem{criuSetPid}
{CRIU team}, ``{CRIU} --- pid restore,'' accessed Jan., 2017,
  \url{https://criu.org/Pid_restore}.

\bibitem{criuPlugins}
------, ``{CRIU} --- plugins,'' accessed Jan., 2017,
  \url{https://criu.org/Plugins}.

\bibitem{ValleeEtAl05}
G.~Vallee, R.~Lottiaux, D.~Margery, and C.~Morin, ``Ghost process: A sound
  basis to implement process duplication, migration and checkpoint/restart in
  {L}inux clusters,'' in \emph{Proc. of the The 4th Int. Symp. on Parallel and
  Distributed Computing}, ser. ISPDC '05, 2005, pp. 97--104.

\bibitem{PorterEtAl11}
D.~E. Porter, S.~Boyd-Wickizer, J.~Howell, R.~Olinsky, and G.~C. Hunt,
  ``Rethinking the {L}ibrary {OS} from the top down,'' in \emph{Proc. of the
  Sixteenth International Conference on Architectural Support for Programming
  Languages and Operating Systems}, ser. ASPLOS XVI.\hskip 1em plus 0.5em minus
  0.4em\relax New York, NY, USA: ACM, 2011, pp. 291--304.

\bibitem{EnglerEtAl95}
D.~R. Engler, M.~F. Kaashoek, and J.~O'Toole, Jr., ``Exokernel: An operating
  system architecture for application-level resource management,'' in
  \emph{Proc. of 15th ACM Symp. on Operating Systems Principles}, ser. SOSP
  '95.\hskip 1em plus 0.5em minus 0.4em\relax ACM, 1995, pp. 251--266.

\end{thebibliography}
\end{document}